\documentclass[twocolumn,secnumarabic,amssymb, nobibnotes, aps, prb,superscriptaddress]{revtex4-1}
\usepackage{amsmath,amssymb,amsfonts}
\usepackage{epsfig}
\usepackage{graphicx}
\usepackage{hyperref}
\usepackage{natbib}
\hypersetup{colorlinks=true,citecolor=blue,linkcolor=blue,urlcolor=blue}
\usepackage[all]{hypcap}
\usepackage{url}
\usepackage{color}

\newcommand{\U}{\textit{U}}
\newcommand{\J}{\textit{J}}

\newcommand{\ohd}{$O_h^{\downarrow}$}
\newcommand{\ohu}{$O_h^{\uparrow}$}
\newcommand{\tdd}{$T_d^{\downarrow}$}
\newcommand{\tdu}{$T_d^{\uparrow}$}
\usepackage{soul}
\begin{document}
\title{\large{Bismuth iron garnet: \textit{ab initio} study of electronic 
properties}}
 \author{\firstname{Federico} \surname{Iori}}
\affiliation{Laboratoire de Physique des Solides, CNRS, Univ. Paris-Sud, 
 Universit\'e Paris-Saclay, 91405 Orsay Cedex, France}
\author{\firstname{Adrien} \surname{Teurtrie}}
\affiliation{Laboratoire de Physique des Solides, CNRS, Univ. Paris-Sud, 
Universit\'e Paris-Saclay, 91405 Orsay Cedex, France}
\affiliation{Groupe d'Etude de la Mati\`ere Condens\'ee (GEMaC), CNRS-UVSQ, Universit\'e 
Paris-Saclay, 78035 Versailles, France}
\author{\firstname{Laura} \surname{Bocher}}
\affiliation{Laboratoire de Physique des Solides, CNRS, Univ. Paris-Sud, 
Universit\'e Paris-Saclay, 91405 Orsay Cedex, France}
\author{\firstname{Elena} \surname{Popova}}
\author{\firstname{Niels} \surname{Keller}}
\affiliation{Groupe d'Etude de la Mati\`ere Condens\'ee (GEMaC), CNRS-UVSQ, Universit\'e 
Paris-Saclay, 78035 Versailles, France}
\author{\firstname{Odile} \surname{St\'ephan}}
\author{\firstname{Alexandre} \surname{Gloter}}
\affiliation{Laboratoire de Physique des Solides, CNRS, Univ. Paris-Sud, 
Universit\'e Paris-Saclay, 91405 Orsay Cedex, France}

\begin{abstract}
Bismuth iron garnet (BIG), i.e. Bi$_3$Fe$_5$O$_{12}$, is a strong ferrimagnet that also 
possess outstanding magneto-optical properties such as the largest known Faraday 
rotation. 
These properties are related with the distribution of magnetic moments on octahedral and 
tetrahedral sites, the presence of spin gaps in the density of state and a strong 
spin-orbit coupling. In this work, first-principles \textit{ab initio} calculations are 
performed to study the structural, electronic and magnetic properties of BIG using 
Density Functional Theory with ”Hubbard+\U” (DFT+\U) correction including spin-orbit 
coupling and HSE06 hybrid functional. We found that the presence of spin gaps in the 
electronic structure results from the interplay between exchange and correlation 
effects and the crystal field strengths for tetrahedral and octahedral iron sublattices.  
The DFT+\U~ treatment tends to close the spin-gaps for larger U due to over-localization 
effects, notably in the octahedral site. On the other hand, the hybrid functional confirms 
the occurrences of three spin gaps in the iron states of the conduction band as expected 
from optical measurements. A strong exchange splitting at the top of the valence bands 
associated with a lone-pair type mixture of O \textit{p} and Bi \textit{s,p} states is 
also obtained. Similar exchange splitting was not previously observed for other iron based 
garnets, such as for yttrium iron garnet. It follows that hole doping, as obtained by Ca 
substitution at Bi sites, results in a full spin polarized density at the Fermi energy. 
This work helps to shed more light on the theoretical comprehension of the properties of 
BIG and opens the route towards the use of advanced Many Body calculations to predict the 
magneto-optical coupling effects in BIG in a direct comparison with the experimental 
measurements.

\end{abstract}
\maketitle
\noindent
\section{Introduction}
Bi$_3$Fe$_{5}$O$_{12}$~(BIG) is a ferrimagnet insulator exhibiting a
magneto-electric coupling at 300 K, as recently reported~\cite{popo+17apl}. Contrary to 
its 
parent structures such as, for example, the well-known yttrium iron 
garnet (YIG)~\cite{popo+01jap}, this material can only be synthesized in thin-film form
using non-equilibrium growth 
techniques~\cite{vert+08prb}. Nevertheless the growth 
effort worths the price since ferrimagnetic bismuth iron garnet shows relatively 
high magnetization of 1.27x10$^5$ A/m at 300 K and magnetic ordering 
temperature from 650 K to 700 K, depending on Bi content and film 
thicknesses~\cite{popo+13jmm,vert+08prb}. Moreover, BIG giant Faraday rotation 
effect makes this material a suitable candidate for fast magneto-optical 
sensors~\cite{gorn+10cgt}, optical isolators~\cite{dred-yosh09optex} and second 
harmonic generation~\cite{prad+10opte}. Despite the significant technological 
interest, the structural and 
electronic properties of BIG thin films remains debated. In literature the preliminary 
paper of 
Oikawa and coworkers~\cite{oika+05jpsj} investigates the electronic structure of BIG 
through an {\it{ab initio}} approach based on spin polarized local density 
functional theory (LSDA) and full-potential 
linear-combination-of 
atomic-orbitals (LCAO) without including any correction treatment for the ''correlated'' 
localized 
3\textit{d} electrons.

As known from 
literature~\cite{iori+12prb,rodl+09prb}, local and semilocal 
functionals suffer of severe delocalization errors~\cite{mori+08prl} 
particularly relevant for localized \textit{d} and \textit{f} electrons. In 
transition metal oxides, this underestimation can lead to the prediction 
of metallic band structure instead of the correct insulating 
one~\cite{iori+12prb}. In the case of BIG this 
leads to a poor description of the electronic charge density and 
underestimation of band gaps and magnetic moments compared to experimental 
evidences~\cite{oika+05jpsj}. Recent magneto-optical 
measurements~\cite{deb+13prb,koen+15prb,vert+08prb} 
re-open the questions regarding the precise description of the spin-dependent 
electronic structure of BIG. The optical 
measurements by Kahl~\cite{kahl+03jap} in 2003 reports an optical absorption 
gap in BIG of 2.3 eV. Recently Popova and 
coworkers~\cite{deb+13prb,popo+17apl} investigated through magneto-optical 
spectroscopy the spin-dependent electronic density of states (DOS)
near and above the Fermi level in BIG. In particular, magneto-optical measurements have 
revealed a strong asymmetric Faraday hysteresis loop for some photon energies that have 
been related to the presence of spin gaps in the conduction bands~\cite{deb+13prb}.

This work attempts to shed more 
light on the theoretical description of the electronic and magnetic properties 
of BIG. In particular, we investigate how the interplay between the treatment of the 
electronic 
correlation and the crystal field for the tetrahedral and octahedral sites can modify the 
spin polarization of the calculated density of states. We 
also evaluate the influence of the spin-orbit coupling in this density of states. 
Furthermore, we report an exchange splitting in the top of the valence bands constituted 
of a mixture of O\textit{p}, Bi\textit{s}, Bi\textit{p} states typical of a lone pair. 
Finally, 
first-principles studies of Ca substituted 
BIG indicate that a full spin-polarized density at the Fermi energy might 
be obtained in BIG by doping. 

\section{Methods}
We have used \textit{ab initio} calculations based on 
Kohn-Sham Density Functional Theory (DFT)~\cite{kohn-sham65pr,hohe-kohn64pr}
in a planewave pseudopotential approach as implemented in {\scshape{Quantum 
Espresso}}~\cite{pwscf}. Scalar and full relativistic ultrasoft pseudopotential, 
including semicore states and spin-orbit coupling for Bi and Fe, have been obtained from 
the PSLibrary designed by DalCorso~\cite{dalc14cms}. An energy cutoff of 50 Ry for 
plane-wave basis expansion and of 400 Ry to describe the charge density and the 
potential have been respectively used. Geometry has been relaxed at the Gamma 
k-point until reaching a maximum force on each atom smaller than 
10$^{-4}$ eV/\AA. Charge density has been converged with 2$\times$2$\times$2 and 
4$\times$4$\times$4 k-points and the density of states with a total mesh of 64
k-points in total. We applied a spin-polarized generalized gradient 
approximation exchange-correlation functional in the Perdew-Burke-Ernzerhof 
(PBE)~\cite{perd+96prl,bart-hedi72jphysc} formulation and, for comparison, in its 
optmized version for solids (PBEsol)~\cite{perd+08prl}.
To improve the treatment of electronic correlation in the description of the 
correlated subset of Fe~\textit{d} states we apply a 
''Hubbard+{\it{U$_{eff}$}}, with {\it{U$_{eff}$=\U-\J}}'' scheme, where 
\textit{U} represent the \textit{ad hoc} Hubbard on-site Coulomb repulsion parameter and 
\J~the Hund's exchange~\cite{liec+95prb}
We used the effective ''Hubbard-\textit{U}''~\cite{duda+98prb} in the simplified 
formulation of Cococcioni 
et.~\cite{coco-degi05prb} and for 
spin-orbit calculation the rotationally invariant approach 
by Liechtestein et al.~\cite{liec+95prb}. We evaluate the effect of several values of 
\U~keeping the onsite Hund's exchange \textit{J}=0, which formally renders the two 
approaches 
equivalent~\cite{baet+05prb} and for the specific case of \U=4 eV we checked the effect 
of the onsite exchange by varying \textit{J}. The \U-corrected functionals will thus be 
referred as PBE+\U~and PBEsol+\U. The hybrid functional have been applied 
within the HSE06~\cite{heyd+03jcp} formulation as implemented in 
the {\scshape{Vasp}}~\cite{vasp} code.

\begin{figure}[h]
\centering
\includegraphics[angle=0,width=1\columnwidth]{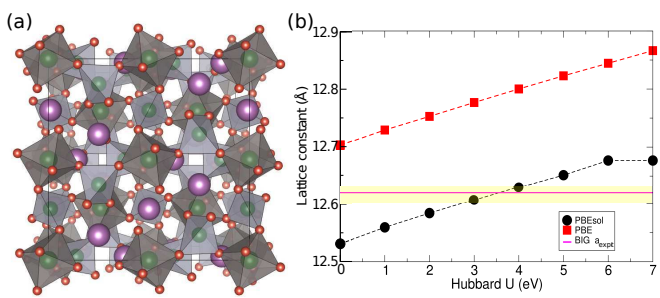}
\caption{(Color online).Panel (a): Conventional crystal structure of bismuth iron garnet 
with 
{\textit{Ia-3d}} symmetry plotted along [001] direction. Tetragonal (light grey) and 
octahedral (dark grey) polyhedra centered on Fe atoms (green) with
O atoms (red) at the corners. Bismuth atoms are in violet. 
Structure plotted with {\scshape{VESTA}}\cite{vesta}. Panel (b):  
Theoretical lattice constants a$_{theo}$ optimized 
with PBE+\U~(red squares) and PBEsol+\U~(black dots) as a function of 
\U~. The yellow bar indicate the range of variation of the experimental
lattice constant and the purple line correspond to the value
a$_{expt}$=12.624~\AA~ used in this work, as explained in the text.}
\label{fig1}
\end{figure}

\section{Influence of the Hubbard correction}
The conventional cell (space 
group {\textit{Ia-3d}}) contains eight chemical formulas for a total of 160 
atoms. As yttrium in Y$_3$Fe$_{5}$O$_{12}$~\cite{chin+01jap,such+13com,gell+63apl},
the 24 Bi cations of BIG occupy dodecahedral coordinate sites, while the Fe cations are in
different coordination sites: 24 tetragonal Fe $T_d$ spin up ($\uparrow$) and 16 
octahedral Fe $O_h$ spin down ($\downarrow$), forming two magnetic 
sublattices ferrimagnetically coupled as shown in Fig.~\ref{fig1}, panel (a).
 
Since the magnetizations of different 
sublattices do not completely cancel each other, a spontaneous ferrimagnetic polarization 
raises up. BIG grows only epitaxially in thin film form requiring isostructural garnet 
substrates i.e.
Y$_3$Al$_5$O$_{12}$ (YAG), Gd$_3$Ga$_5$O$_{12}$ (GGG) or substituted GGG 
(SGGG) having smaller lattice constants than of BIG (12~\AA~(YAG), 12.37~\AA~(GGG) and
12.48~\AA~(SGGG) respectively~\cite{kahl+02jap,popo+17jap}) that result in 
compressive strain to the BIG film. Therefore the measurement of the absolute 
value of the lattice constant remains particularly challenging~\cite{popo+17jap}.
In this context, the experimental lattice constant of BIG thin films varies according to 
literature between 12.60 and 12.633~\AA~\cite{ada+00jap,okad+91jap,okud+90jap}, while 
polycrystalline~\cite{tora-okud95jpcs} and 
monocrystalline~\cite{mino+92jjap} BIG films have reported to present 
slightly larger values of 12.64 or 
12.65~\AA. We calculated the theoretical lattice $a_{theo}$ constant optimized with 
PBE+\U~and PBEsol+\U~for \U~ranging from 0 to 7 eV~and we reported them for comparison in 
Fig.~\ref{fig1} panel (b).

\begin{figure}[th]
\centering
\includegraphics[angle=0,width=1\columnwidth]{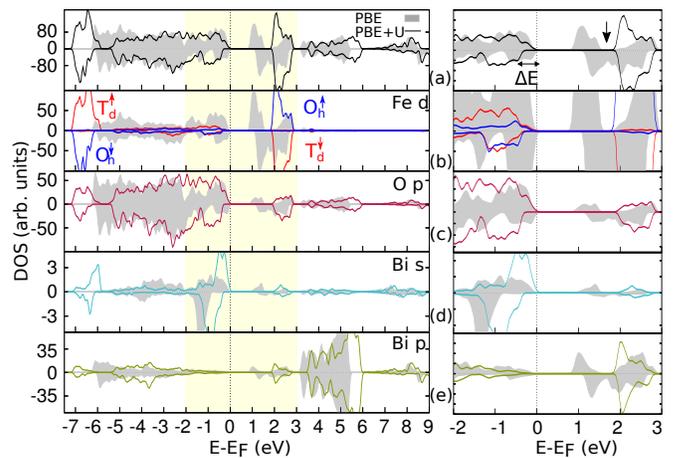}
\caption{(Color online.) {Total DOS (TDOS) and projected density of states (PDOS) of BIG 
for PBE in grey filled curve and PBE+\U=4 eV in solid line. Left panel shows the overall 
TDOS and PDOS. Right panel shows the zoom from -2 to 3 eV corresponding to the 
highlighted part in yellow in the left panel.
Panel~(a). DOS; PBE in grey, PBE+\U~in solid line. 
Panel~(b). PDOS of Fe~\textit{d} states: blue curve corresponds to
octahedral Fe \ohd~(filled) and \ohu~(empty);
red curve corresponds to the tetrahedral Fe \tdu~(filled) and \tdd~(empty).
Panel(c). PDOS of O~\textit{p} states. Panel(d).
PDOS of Bi \textit{s} states, intensity $\times$ 5. Panel(e).
PDOS of Bi \textit{p} states.}}
\label{fig2}
\end{figure}

From Fig.~\ref{fig1} panel (b) it emerges 
that: 
(i) the PBE and PBEsol theoretical optimized BIG lattice parameters differ from the 
experimental lattice 
constant of the garnet substrates by an amount of 1\% to 5\%, as experimentally 
observed~\cite{popo+17jap} (ii) the theoretical lattice constants optimized with 
PBEsol+\U~for \U among 2 and 4.5 eV fall into the experimental range of variation 
indicated by the yellow bar in Fig.~\ref{fig1}. An absolute calculation 
of the BIG lattice constant strongly depends on the 
exchange-correlation functional used and on the theoretical level of approximation
applied to treat the strong electronic correlation effects. To that, hereafter we 
will choose for our calculations a lattice 
constant equal to an average value $a_{expt}$ of 12.624~\AA, a choice in accordance with 
previous theoretical studies~\cite{oika+05jpsj} and with the PBEsol calculation for~\U=4 
eV. Therefore, we 
will take~\U=4~eV as the 
reference value for the Hubbard 
correction. This choice is also coherent with recent 
first-principles DFT+\U~studies on rare-earth 
ferrites~\cite{dieg+11prb,xu+14prb,ching+01jap,neat+05prb} and
garnets~\cite{naka+17prb} where it has been shown that DTF+\U=4 eV for such 
oxides permits to achieve a whole qualitatively and quantitatively consistent description 
of the structural and electronic properties~\cite{shen+17jpc}.
\begin{table}[h]
\begin{ruledtabular}
\caption{{\label{tab1}}{Electronic and magnetic properties 
calculated within PBE+\U~and PBEsol+\U~for \U=0 and \U=4 eV using the average 
experimental lattice parameter ($a_{expt}$=12.624 \AA)  The first three rows show the 
direct band gap ($E_{D}$), 
the band gap ($E_{\uparrow}$) for the spin-up channel and the band gap 
($E_{\downarrow}$) for the spin-down in eV and calculated at 
$\Gamma$ 
k-point. The last six rows report the total and iron magnetic moment per 
formula unit 
($\bar{\mu}_{tot}$, $\bar{\mu}_{Fe}$) and the moments per atom of each 
chemical 
species in units of $\mu_{B}$.}}
\setlength{\tabcolsep}{3pt}
\begin{tabular}{l c c c c c c}
  & \multicolumn{2}{c}{PBE ($a_{expt}$) } & \multicolumn{2}{c}{PBEsol ($a_{expt}$) } & 
Literature\\
\cline{2-3} \cline{4-5}
  &  \U=0 & \U=4 & \U=0 & \U=4& \\  
\hline
\hline
E$_{D}$ 		& 0.86 & 1.90  & 0.77  & 1.81 	&2.3\footnotemark[1]\\
E$_{\uparrow}$ 		& 0.91 & 1.90  & 0.83  & 1.81 	&2.3\footnotemark[1]\\
E$_{\downarrow}$ 	& 1.17 & 2.30  & 1.11  & 2.23 	&2.3\footnotemark[1]\\
$\bar{\mu}_{tot}$	& 3.96 & 4.00 & 3.94  & 3.98  & 4.25-5.0\footnotemark[2]\\
${\bar\mu}_{Fe}$	& 2.94 & 3.09 & 2.92  & 3.07  & 3.45\footnotemark[3]\\
$\mu_{Fe(o)}$ 	  	& -3.51& -3.81& -3.46 & -3.80 &  -3.27\footnotemark[3]\\
$\mu_{Fe(t)}$	  	& 3.32 & 3.57 & 3.28 & 3.55   &  3.94\footnotemark[3]\\
$\mu_{O}$	  	& 0.08 & 0.07 & 0.08  & 0.07  &  0.106\footnotemark[3]\\
$\mu_{Bi}$	  	& 0.002& 0.005& 0.002 & 0.005 &  -
\label{table1}
\end{tabular}
\end{ruledtabular}
\footnotetext[1]{Experimental optical gap from Ref.~\cite{kahl+03jap}}
\footnotetext[2]{Experimental value for YIG from Ref.~\cite{pasc+84prb,rodi+99jmm}}
\footnotetext[3]{Theoretical values for YIG from Ref.~\cite{chin+01jap}}
\end{table}
The main fingerprints of the electronic structure of BIG are then reported in 
Table~\ref{table1} and Fig.~\ref{fig2}. Table~\ref{table1} contains the $\Gamma$ 
direct electronic gaps~\cite{gapnote} 
and the magnetic moments calculated for both PBE and PBEsol functionals and 
different Hubbard-\U (\U=0 and \U=4 eV) using $a_{expt}$ lattice 
constant.

For the same values of \U, there are no remarkable 
differences between the electronic gaps and the magnetic moments obtained by using PBE 
or PBEsol as shown in Table~\ref{table1}. Nonetheless by applying \U=0 to 4 eV for 
both functionals, we get a significant 
increase of 
the direct electronic gap E$_{D}$ (E$_{D}$ almost doubles to 1.90 eV in PBE+\U~and 1.81 
eV in 
PBEsol+\U~for $a_{expt}$) and a slight enhancement of the magnetic moments. Therefore in 
our case the change of the electronic gaps and magnetic 
moments is related to the approximation chosen to treat the electronic correlation rather
than to the impact of the lattice parameters. PBE+\U~and PBEsol+\U~give very close 
numerical results. The effects of the introduction of 
electronic correlation are then depicted in Fig.~\ref{fig2} where are shown the 
electronic total density of states (TDOS) and the orbital-projected density of 
states (PDOS) calculated 
within PBE (grey) and PBE+\U~(coloured) approximation.
With respect to Fig.~\ref{fig2} PBE reproduces an insulating density of states with 
spectral weight in general agreement with the LSDA calculation 
by Oikawa~\cite{oika+05jpsj} but with larger electronic band gaps. The $\Gamma$ direct 
gap is 0.86 eV, 0.91 eV for spin-up and 1.17 eV for 
 spin-down channel (as shown in Table~\ref{table1}). 

 The PBE projected density of states shows a valence band ranging from 0 to -6 eV
characterized by a mixing of O 2\textit{p}, Bi 6\textit{s} and Fe 3\textit{d}.
The PBE conduction states are 
dominated by the spin-polarized Fe 3\textit{d} in agreement to previous LSDA 
calculation~\cite{oika+05jpsj}. The 
ferrimagnetism in BIG arises clearly from the different spin polarization intensity 
among the
Fe 3\textit{d} electrons in the two Fe $T_d$ and Fe $O_h$ sublattices as shown in
Fig.~\ref{fig2} 
panel(a). Moreover the crystal-field effects split the Fe 3\textit{d} in a 
triple degenerate $t_{2g}$ states and in doubly degenerate $e_{g}$ 
states with different energetic hierarchy for both sublattices. The clear 
separation of 
Fe 3\textit{d} bands for the spin-up conduction band is due to a 
stronger $O_h$ crystal-field. For the minority spin, instead, the 
$T_d$ crystal-field is not strong enough to separate the 
$e_{g}/t_{2g}$ contribution in the unoccupied part. This strong difference of crystal 
field strengths for $O_h$ and $T_d$ symmetry was also experimentally reported in the case 
of YIG based on the analysis of the d-d optical transitions where a crystal field 
splitting of 1.52 eV and 0.77 eV were derived for respectively $O_h$ and $T_d$ 
sites~\cite{gavr+05jetpl}. The geometry of the BIG garnet 
(minority spin in pure O$_h$ site and majority spin in pure $T_d$ site for the occupied 
band) is at the origin of the 100\% spin polarized density of states that occurs 
at several energies in the DOS calculated by PBE. Indeed the spin-up conduction 
\ohu~states are forming two separated peaks centered at 1.2 and 3 eV while the spin-down 
\tdd~states have a unique peaks at around 2 eV as reported in Fig.~\ref{fig2} panel 
(a,b) by the grey filled curves.

The introduction of the on-site Coulomb interaction 
Hubbard-\U~over PBE gives rise to an energy splitting between occupied and empty states
in such a way that the former are pushed down and the latter up in energy, as 
seen in literature for other transition 
metal oxides and rare-earth garnet~\cite{rodl+09prb,naka+17prb}. In Fig.~\ref{fig2}, in 
coloured curves, is 
shown such PBE+\U~opening of the band gap followed by the 
redistribution of 
spectral weight of the density of states. The O 2\textit{p} states extend 
from the Fermi energy to ${\sim}$ -7 eV but the PBE+\U~helps to break the 
hybridization in valence band separating
the O 2\textit{p} from the Fe 3\textit{d} states but still showing a non negligible 
superexchange character. The top of 
valence band around the Fermi level is dominated by 
the O 2\textit{p} hybridized with Bi 6\textit{s} states. The ferrimagnetic exchange 
splitting $\Delta$E between spin up and spin down state at the top of the valence 
band
(shown in 
panel (a) of Fig.~\ref{fig2}) is clearly enhanced for~\U=4 eV and correspond to 0.36 eV. 
The interaction between the Bi \textit{s,p} and the O \textit{p} orbitals
corresponds in our case exactly to the typical ''lone pair model'' seen in other 
oxides~\cite{wals+07chemmat,wals+11chesocrev}. Indeed, the Bi atoms sit in 
non-centrosymmetric distorted dodecahedra allowing the mixing of on-site Bi \textit{s} 
and 
Bi \textit{p} 
orbitals typical of the lone pair mechanism. According to this model, strong interaction 
between the Bi \textit{s} and O \textit{p} determines the separation in bonding and 
anti-bonding states at high energy. This results in the strong contribution of an 
antibonding Bi \textit{s} orbital character at the top 
of the upper valence band.

\begin{table}
\begin{ruledtabular}
\caption{PBE+\U~electronic gaps and
magnetic moments as a function of~\U. The total band gap E$_g$ is calculated as the 
energy difference between the lowest energy conduction 
state and the highest energy valence state at the $\Gamma$ k-point; A,B,C are the 
exchange gaps in the conduction density of states as shown in 
 Fig.~\ref{fig3}. The magnetic moments $\mu$ are calculated per formula unit (FU).} 
\begin{tabular}{l c c c c c c}
 U & E$_{g}$ & A & B & C & $\mu$/FU 
& $\mu_{Fe}$/FU\\
  & (eV)      & (eV)       & (eV)       & (eV)       & ($\mu_B$)  & ($\mu_B$)\\
\hline
\hline
 0 & 0.91 & 0.31 & 0.20 & 0.25 & 3.96  & 2.40\\
 1 & 1.19 & 0.29 & 0.10 & 0.25 & 3.97  & 3.98\\
 2 & 1.42 & 0.27 & -    & 0.15 & 3.98  & 3.01\\
 3 & 1.66 & 0.23 & -    & 0.24 & 3.98  & 3.04\\
 4 & 1.90 & 0.05 & -    & -    & 4.00  & 3.09\\
 5 & 1.91 & 0.07 & -    & -    & 4.00  & 3.15\\
 6 & 1.77 & 0.11 & -    & -    & 4.02  & 3.2
\label{tab2}
\end{tabular}
\end{ruledtabular}
\end{table}
The bottom of conduction has a Fe \textit{d} character while the Bi \textit{s} and 
\textit{p} states are upward shifted in energy of almost 1 eV with respect to the PBE 
calculation. The O 2\textit{p} and Bi 6\textit{p} augment their mutual hybridization 
between 3.5 eV and 10 eV. The effect of \U~on the conduction states is more pronounced 
for the Fe \textit{d}.
The inclusion of Hubbard \U~on the 
Fe $T_d$~and Fe $O_h$~states tends to shrink their bandwidth from ~2 to ~1.2 eV. The 
effect is 
more evident for the $O_h^{\uparrow}$ states where the $t_{2g}-e_g$ splitting (indicated 
for the U=0 calculation by the black arrow in Fig.~\ref{fig2}) disappears.

\begin{figure}[h]
\centering
\includegraphics[angle=0,width=1\columnwidth]{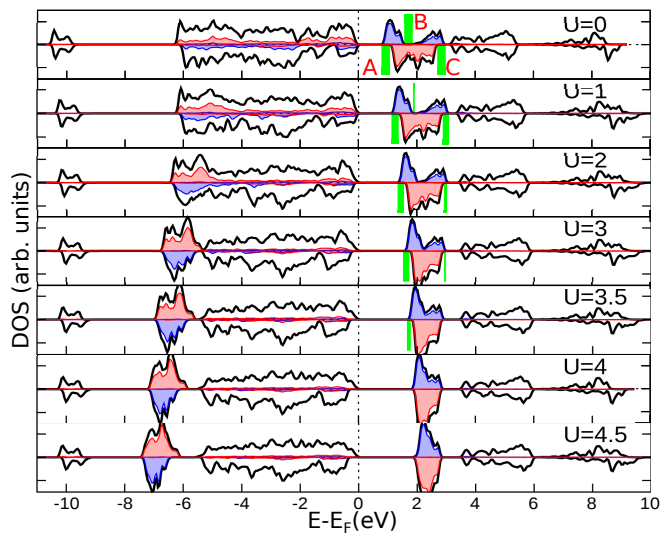}
\caption{(Color online.){Evolution of the Fe 
3\textit{d} states (Fe $O_h$ in blue and Fe $T_d$ in red) and of the exchange splitting 
gaps 
(A,B,C - green lines) among the $O_h^{\uparrow}$ and the $T_d^{\downarrow}$ \textit{d} 
empty states for Hubbard-\U~ranging from 0 eV to 4.5 eV. Their values are 
reported in Table~\ref{tab2}. The total DOS is reported as reference in 
solid black line for each \U.}}
\label{fig3}
\end{figure}

In order to interpret the spin gap closure of the $O_h^{\uparrow}$ 
states we have evaluated the effects of Hubbard \U~on the $O_h$ and $T_d$ crystal field 
splitting for \U~ranging from 0 eV to 4.5 eV as shown 
in Fig.~\ref{fig3} and reported in Table~\ref{tab2}. For 
increasing value of 
\U~the 3\textit{d} states change completely their energy position. In valence band the Fe 
states spectral weight is depleted from the Fermi level and shifted downward
in energy with an important reduction of the hybridization between the O and Fe orbitals. 
For \U=4 eV, the Fe majority occupied states $T_d^{\uparrow}$ and the 
minority spin  
$O_h^{\downarrow}$ form two antiferrimagnetic peak centered around 
-7 eV of Fermi Energy (FE) that become separated from the main oxygen band. With 
increasing the value of \U, it is noteworthy that the top of the valence band also 
becomes spin split.

\begin{figure}[h]
\centering
\includegraphics[angle=0,width=1\columnwidth]{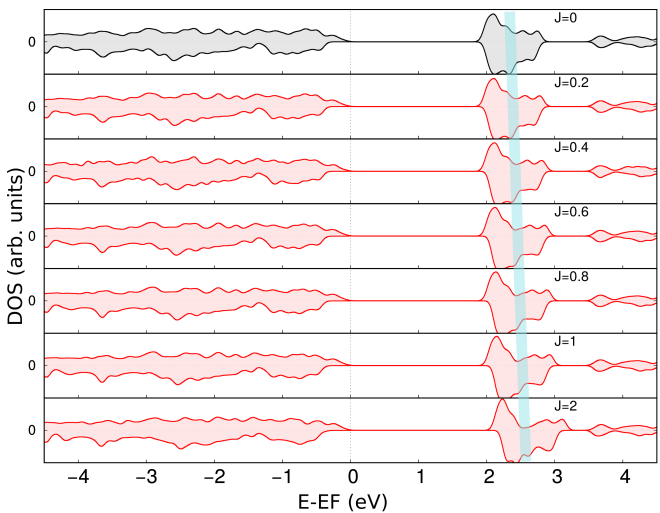}
\caption{(Color online.) Effects of Hund's exchange \J~on the PBE+\U-\J~density of 
states of BIG for \U=4 eV, \J~varying from 0 to 2 eV. In topmost panel 
(in grey) the calculation of DOS for \J=0, corresponding to the PBE+\U DOS shown in 
Fig.~\ref{fig2}. The highlighted cyan window shows the increase of the exchange 
splitting for \J$>$0.}
\label{fig4}
\end{figure}

In the conduction band the principal effect of \U~is to reduce the bandwith of both 
spin-up 
and spin-down Fe 3\textit{d}. The main effect is the 
disappearance of the gap between the spin-up $O_h$ states present in the PBE 
calculation (indicated by the black arrow in Fig.~\ref{fig2} 
and B in Fig.~\ref{fig3}).
%
%
This closure is linear with 
\U~and it appears clearly at \U=3 eV with a transition in DOS from a structure 
with two peaks to a structure with only one well-defined peak. 
The localization effects of the Hubbard term is stronger for the spin-up composed 
of $O_h$ symmetry with larger hybridization than for the spin-down with pure $T_d$ 
contribution. At moderate (for a transition metal oxide) value of \U, the spin gaps 
observed at the three 
energies A,B,C (reported in Fig.~\ref{fig3} in the DOS tend 
to disappear, while 
magneto-optic measurements have revealed the existence of several spin 
gaps~\cite{deb+13prb}. On the 
contrary the total 
and the Fe magnetic moments per formula units are almost stabilized for 
different \U~values around $\mu=3 \mu_B$ and $\mu=4 \mu_B$ respectively.
The Hubbard term therefore aids to correct the delocalization error of pure 
DFT-PBE, but seems to induce a spurious artifact in the correct description of the 
hybridization between oxygen and the iron states~\cite{shen+17jpc}.
We then take into account the effect of the Hund's exchange \J~by performing a fully 
rotationally invariant PBE+{\it{U$_{eff}$}},
calculations~\cite{kesh+18prb}. For fixed \U=4 eV we varied \J~and monitored 
the evolution of electronic and magnetic properties. In Fig.~\ref{fig4} is reported the 
evolution of the total DOS for PBE+{\it{U$_{eff}$}}, for \J~ranging from 0 to 2 eV. 
We can see that for \J$>$0, the exchange splitting among the conduction Fe {\it{d}} 
states increases progressively (cyan highlighted region in Fig.\ref{fig4}) as observed in 
similar work in literature~\cite{chen-mill16prb}. The electronic $\Gamma$ direct gap is 
increasing between 1.90 eV for \J=0 and 2 eV for \J=2 eV while the Fe magnetic moments 
$\mu_{Fe(o)}$ and $\mu_{Fe(t)}$ are stabilized around -4 $\mu_B$ and 3.8$\mu_B$ 
respectively. Interestingly for \J=2 eV the huge exchange splitting among the Fe {\it{d}} 
counteracts the over localization observed for PBE+\U, \U=4 eV permitting to recover the 
octahedral crystal field splitting and an almost full spin polarization behaviour in the 
conduction states.

\begin{figure}[t]
\centering
\includegraphics[angle=0,width=1\columnwidth]{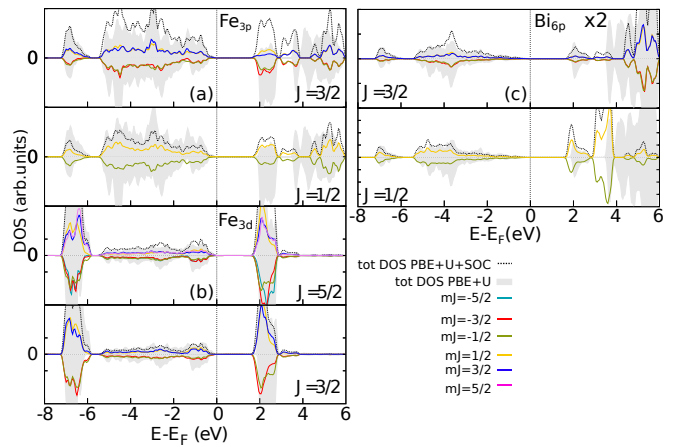}
\caption{(Color online.){The PDOS PBE+\U+SOC and 
and PBE+\U spin polarized cases. In panel (a) the Fe 
3\textit{p}, in panel (b) the Fe 3\textit{d}, in panel (c) the Bi 6\textit{p}. 
In each panel PBE+\U DOS is in grey, the total PBE+\U+SOC DOS is in dotted black line.
The total $l$-resolved DOS per element (dotted black line) is the sum of all the 
corresponding m$_J$ projections (solid colored lines). The intensity of 
the Bi 6\textit{p} has been magnified by 2.}}
\label{fig5}
\end{figure}

Moreover the role of the spin-orbit (SO) coupling of the bismuth and iron atoms on 
the 
density of states of the BIG has been evaluated. 
In Fig.~\ref{fig5} are compared the atom-projected PBE+\U~density of states with the 
PBE+\U+SOC density of states projected for different value of the moment $m_J$.  
The SOC affects 
mainly the Bi 6\textit{p} and only slightly the Fe 3\textit{p} and Fe 3\textit{d}.

As compared to the PBE+\U~calculations where the Bi \textit{p} main states were spanning 
from 3 to 
6 
eV, the PBE+\U+SOC calculated DOS results in a Bi (6\textit{p}$_{1/2}$) spanning from 2 
to 4 
eV and the Bi (6\textit{p}$_{3/2}$) from 4 to 7 eV. The Fe 3\textit{p} and Fe 3\textit{d} 
state are then 
more hybridized with the bismuth when SOC is included via the Bi 
(6\textit{p}$_{1/2}$) states. The SO effect on iron \textit{d} states is instead very 
soft. In the Fe 3\textit{d} 
valence band the SO splitting is 50 meV,  while in conduction is around 25 meV, in good 
agreement with experimental and calculated values reported by Oikawa~\cite{oika+05jpsj}.

%

\section{Hybrids functional applied to BIG}
We have also calculated the electronic structure with
HSE06 hybrid functional on top of PBE as implemented in 
{\scshape{Vasp}}~\cite{vasp}. HSE06 improves the 
electronic structure description and the correct orbital occupation beyond a local or 
semi-local approximation stemming from treating
all the electrons on the same footing~\cite{rodl+09prb,fran14jpc,choi+13prb}.
This hybrid functional can be considered one of the best static non-local approach to 
GW approximation in order to reproduce band gap and optical dielectric functions close to 
experiments~\cite{marq+11prb,iori+12prb}.
%
\begin{figure}[h]
\centering
\includegraphics[angle=0,width=1\columnwidth]{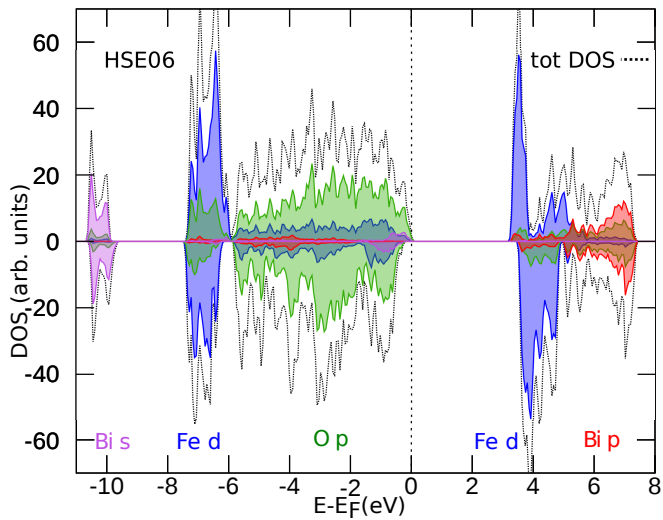}
\caption{(Color online.) HSE06 hybrid functional calculation of the TDOS and PDOS.
The different projected angular moment channels shown correspond to: Bi 6\textit{s} in 
lilla, Fe 3\textit{d} in blue, O 2\textit{p} in green, Bi 6\textit{p} in red. For 
sake of computational cost 
we have adopted here the 80-atoms rhombohedral primitive cell of BIG 
with a$_{expt}$ lattice constant~\cite{vaspnote}.}
\label{fig6}
\end{figure}
For instance, hybrid functional applied to iron oxides have successfully reproduced the 
electronic structure of LaFeO$_3$~\cite{he-fran21prb}. In particular, the 
energy separation of unoccupied t$_{2g}$ and e$_g$ states have been demonstrated to be 
almost 
independent to the amount of Hartree-Fock exchange and to be in good agreement with X-ray 
absorption spectroscopy and GW calculation. 

The use of PBE-HSE06 hybrid functional is 
then 
expected to be more accurate for the calculation of crystal field split bands.
The PBE-HSE06 total and projected DOS of BIG are shown in Fig.~\ref{fig6}. Interestingly 
the hybrid functional preserves the strong splitting of the octahedral crystal field
and thus the full spin polarized regions of the unoccupied states.

Moreover as seen for other oxides, the non-local 
Fock exchange part contained in the HSE06 increases the band gap by an upward energy 
shift of the empty states with respect to 
PBE+\U as shown in Table~\ref{tab3}: the HSE06 band gap is now 3.23 eV against the 2.04 
eV of PBE+\U. However the effect of the hybrid functional is very similar to the 
PBE+\U-\J~calculation shown in Fig.~\ref{fig4}. Both approaches increases the exchange 
splitting in the conduction {\it{d}} states, stabilizing the Fe magnetic moments around 
the experimental values and opening the electronic band gap closer to more reasonable 
values. To that we can say that once more the exchange effects have to be seriously 
considered too when dealing with correlated oxides.

HSE06 hybrid functional is known to generally overestimate the 
band gap~\cite{iori+12prb}. The HSE06 gap of 3.23 eV can be then considered as a superior 
limit for the electronic band gap value of BIG. The presence of a strong exciton in 
optical spectra would therefore reduce the electronic gap below the value of 3.23 eV. 
This is therefore compatible and coherent with the experimental optical gap of 2.3 
eV measured by Kahl~\cite{kahl+03jap} and then confirmed recently by 
Deb~\cite{deb+13prb}. Hence in absence of photo-emission measurements we can
infer that the real electronic band gap of BIG should be included between 2 and 3 eV for 
a crystal structure associated with $a_{expt}$=12.624 \AA. Moreover HSE06 confirms a 
substantial exchange 
splitting at the top part of the valence band and at -10 eV of the Bi \textit{s} states 
as we 
found for the PBE+\U~calculations. The total and irons magnetic moments per FU remains 
also similar to the values obtained by PBE+\U. The use of the hybrid functional gives 
similar results as the PBE+\U-\J~calculation shown in Fig.~\ref{fig4}. Both approaches 
increases the exchange splitting in the conduction \textit{d} states, stabilizing the Fe 
magnetic moments around the experimental values and opening the electronic band gap closer 
to more reasonable values. The inclusion of a substantial amount of exchange interaction
is then primordial to obtain the correct ground state in these iron garnets. 

\begin{table}
\begin{ruledtabular}
\caption{PBE-HSE06 and PBE+\U~band gaps and
 magnetic moments calculated with {\scshape{Vasp}} 
with $a_{expt}$. E$_g$ is the total band gap direct at $\Gamma$ k-point.} 
\begin{tabular}{l c c c c}
& E$_{g}$ &  $\mu$/FU & $\mu_{Fe}$/FU\\
& (eV)    & ($\mu_B$)  & ($\mu_B$)\\
\hline
\hline
U=0	&0.85&	4.31&	3.34\\
U=4	&2.04&	4.47&	3.68\\
U=5	&2.32&	4.50&	3.75\\
U=6	&2.53&	4.54&	3.83\\
U=7	&2.67&	4.59&	3.91\\
U=8	&2.77&	4.60&	3.98\\
HSE06 	&3.23&	4.48&	3.64
\label{tab3}
\end{tabular}
\end{ruledtabular}
\end{table}
The relatively strong exchange splitting of the top of the valence band is a key figure 
of 
the BIG electronic structure and is not present in the calculated YIG electronic 
structure~\cite{xu+00jap}. The top of the valence band is mostly composed of Bi 
\textit{s} 
and O \textit{p} and such spin-splitting might be of importance for magneto-optic 
excitations where \textit{p-d} transitions are allowed or for electric transport to 
obtain 
an high spin polarization in presence of impurities.

\begin{figure}[h]
\centering
\includegraphics[angle=0,width=1\columnwidth]{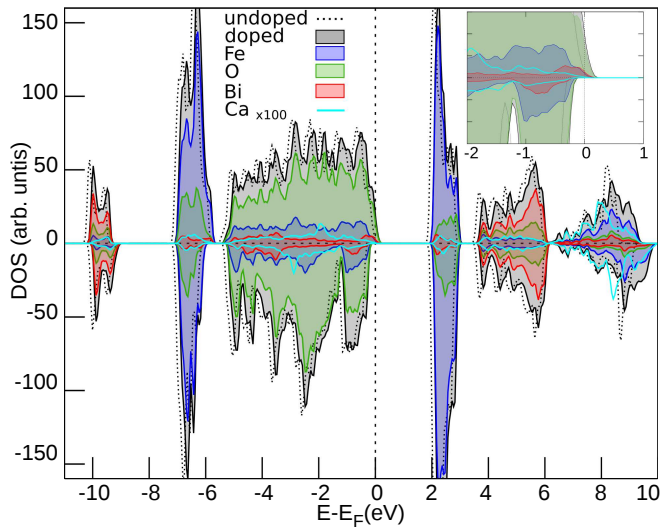}
\caption{(Color online.) PBE+\U (\U=4 eV) total and projected DOS of BIG Ca-doped 
calculated with {\scshape{Quantum Espresso}}. 
(Ca projected DOS has been magnified by 
100). The dotted black line corresponds to the undoped pristine BIG total 
DOS, while the Ca-doped total DOS is in grey. The inset shows a magnification around the 
Fermi energy between -2 and 1 eV.}
\label{fig7}
\end{figure}
In the specific case of hole doping, this could lead to a Fermi 
level with a strong spin polarization. 
We performed calculations with a calcium substituted to a Bi atom corresponding to a 
doping concentration of 4\%. In Fig.~\ref{fig7} are reported the total and projected DOS 
of undoped and Ca-doped BIG calculated within PBE+\U. The calcium atom, stabilized in 
its divalent state, gives rise to an hole doping and an almost 100\% spin polarized 
density of states observed at the Fermi level. At that concentration, no impurity band is 
formed in the band-gap and the non-zero density of states crossing 
the Fermi level indicates that Ca-doped BIG is now a metal due to the hole doping. The 
character of the band at the Fermi level are mostly made of O \textit{p} orbitals for 
such concentration. Nonetheless the Ca-doped BIG appears then as a possible candidate for 
obtaining  a full spin polarized ligand hole electronic structure. 

\section{Conclusion}
To conclude, the occurrence of spin gaps in the Fe 3\textit{d} states of the conduction 
band due to the difference of crystal field value for O$_h$ and T$_d$ sites is observed 
by PBE, 
PBE+\U~for a limited range of \U~ values and by the hybrid functional HSE06 conforming 
the presence of spin gaps as derived from optical measurements~\cite{deb+13prb}. The 
disappearance of the spin gaps at large \U~values is certainly due to some 
over-localization effect of the PBE+\U~approach that mostly alter the O$_h$ sites. The 
inclusion of spin orbit coupling 
mostly changes the unoccupied DOS resulting at lower energy of a Bi \textit{p} (j=1/2) 
hybridized 
with Fe states and at higher energy of Bi \textit{p}(j=3/2) states. The upper part of 
the 
valence 
band is composed of O \textit{p} and Bi \textit{s} states and has also a strong spin 
polarization that was not reported for YIG. The band 
gap obtained by HSE06 gives an E$_g$ of $\sim$ 3.2 eV that can be considered an upper 
limit 
reference for the electronic and optical one. For future outlooks, this work helps to 
shed more light on the 
theoretical comprehension of the properties of BIG and opens the route towards the use 
of advanced Many Body calculations~\cite{sang+12prb} to correctly predict the 
magneto-optical coupling effects in BIG in a direct comparison with the experimental 
measurements.

\section{Acknowledgments}
F.I. acknowledges this research was supported by the Marie 
Sklodowska-Curie IF fellowship (Grant agreement No. 660684) under the European Union 
Horizon 2020 Research and Innovation program. This research was supported  
by ANR Contract DILUMAGOX 12-IS10-0002-01.Computational time was granted by the 
parallel 652-core Linux cluster at Laboratoire de Physique des Solides and by 
GENCI at IDRIS-CNRS Project No. A0020910011. F.I and A.G. are greteful to Alberto 
Zobelli for fruitful discussions.
\bibliography{biblio}

\begin{thebibliography}{60}
\bibitem [{gap()}]{gapnote}%
  \BibitemOpen
  \href@noop {} {}\bibinfo {note} {Our band structure calculated with and
  without SOC (here not shown) give evidence of a $\Gamma$ direct band gap in
  agreement with the few DFT calculations from
  literature~\cite{oika+05jpsj}.}\BibitemShut {Stop}%
\end{thebibliography}
\end{document}